\makeatletter
\declare@file@substitution{revtex4-1.cls}{revtex4-2.cls}
\makeatother

\documentclass[twocolumn]{aastex631}
\usepackage{amsmath, amssymb, graphicx, xcolor}
\usepackage{commath}
\usepackage{hyperref}

\newcommand{\ECHO}{\ensuremath{\tt ECHO}}
\newcommand{\cmbbh}{{\tt CMB-Bh$\overline{a}$rat}}

\begin{document}

%% Title
\title{A perceptron based ILC method to obtain accurate CMB B-mode angular power spectrum}

%% Authors
\author{Sarvesh Kumar Yadav}
\affiliation{Raman Research Institute, CV Raman Avenue, Sadashivanagar, Bangalore, 560080, India}

%% Abstract
\begin{abstract}
Observations of the Cosmic Microwave Background (CMB) radiation have made significant contributions to our understanding of cosmology. While temperature observations of the CMB have greatly advanced our knowledge, the next frontier lies in detecting the elusive B-modes and obtaining precise reconstructions of the CMB's polarized signal in general. In anticipation of proposed and upcoming CMB polarization missions, this study introduces a novel method for accurately determining the angular power spectrum of CMB B-modes. We have developed a Neural Network-based approach to enhance the performance of the Internal Linear Combination (ILC) technique. Our method is applied to the frequency channels of the proposed ECHO (Exploring Cosmic History and Origins) mission and its performance is rigorously assessed. Our findings demonstrate the method's efficiency in achieving precise reconstructions of CMB B-mode angular power spectra, with errors constrained primarily by cosmic variance. 
\end{abstract}

%% Keywords

%% Introduction
\section{Introduction}
The Cosmic Microwave Background (CMB) serves as a direct remnant from the early universe, providing invaluable information about the inflationary epoch \cite{1981PhRvD..23..347G,1982PhLB..108..389L} and the origins of primordial fluctuations that shaped the cosmos. Advances in CMB observations, particularly of temperature and polarization anisotropies, have significantly contributed to the standard cosmological model, $\Lambda$CDM \cite{2013ApJS..208...20B,2020A&A...641A...6P}. One of the most important goals in modern cosmology is detecting the B-mode polarization of the CMB, which encodes information about primordial gravitational waves generated during inflation \cite{1985SvA....29..607P,1997PhRvL..78.2054S,PhysRevD.55.7368}. Such a discovery would not only offer robust evidence for inflation but also enable us to probe the high-energy physics of the early universe, beyond the capabilities of terrestrial high energy physics experiments. The detection of gravitational waves from inflation has profound implications for fundamental physics, motivating numerous ground-based \cite{2019JCAP...02..056A,2017arXiv171003047L,2019arXiv190704473A} and space based \cite{10.1093/ptep/ptac150} observatories. 

%%One such proposed space mission is the CMB Bharat/ECHO (cite). It is a next-generation project to achieve high-precision measurements of CMB polarization. ECHO is specifically tailored to overcome the limitations of previous missions, providing a comprehensive frequency coverage and higher sensitivity to better distinguish between primordial B-modes and those induced by gravitational lensing. 

The CMB B-mode signal is expected to be weak compared to foreground emissions, given current observational constraints \cite{PhysRevLett.127.151301}. Primordial B-modes are further obscured by lensing-induced B-modes generated by large-scale structures \cite{PhysRevLett.89.011303,PhysRevLett.89.011304,PhysRevD.69.043005}. This astrophysical contamination presents a significant challenge for detecting primordial gravitational waves (PGWs), even in experiments with ultra-low noise sensitivity. Unlike the CMB temperature signal, foreground contributions to the polarized microwave sky are dominated by dust emissions at high frequencies and synchrotron emissions at lower frequencies, with no frequency range where the CMB polarization signal clearly dominates. Accurately obtaining the primordial B-mode signal thus requires an effective disentangling of these polarized Galactic foregrounds.

In recent years, numerous studies have focused on minimizing foreground contamination in the CMB B-mode sky (e.g., \cite{2004MNRAS.354...55B,2009A&A...503..691B,2009AIPC.1141..222D,2011MNRAS.414..615B,2011ApJ...737...78K,2012MNRAS.424.1914A,2016JCAP...03..052E,2018JCAP...04..023R,2016MNRAS.458.2032R,2017MNRAS.468.4408H,2021ApJ...914..119Y}). In this article, we propose a novel component separation methodology that combines Internal Linear Combination (ILC) techniques \cite{2007astro.ph..2198D,Saha:2007gf,2009A&A...493..835D} with Artificial Neural Networks (ANN) \cite{Bishop} to accurately extract the B-mode signal from noise and foreground contaminated signal. By leveraging the machine learning technique, this approach aims to significantly improve the accuracy of CMB B-mode angular power spectrum. The investigation is a step in direction to develop a hybrid method based on established ILC method in CMB data analysis and fast developing field of machine learning to improve and enhance data analysis methods for next generation CMB observatories. 

The remainder of this paper is organized as follows: In Section \ref{sec:formalism}, we describe the formalism underpinning our methodology, including the ILC method, associated biases, artificial neural networks, and the simulations used in our study. Section \ref{sec:results} presents and discusses the results of our approach. Finally, in Section \ref{sec:concl}, we provide conclusions and implications for future CMB studies.

\begin{figure*}[ht]
    \centering
    \includegraphics[width=.8\textwidth]{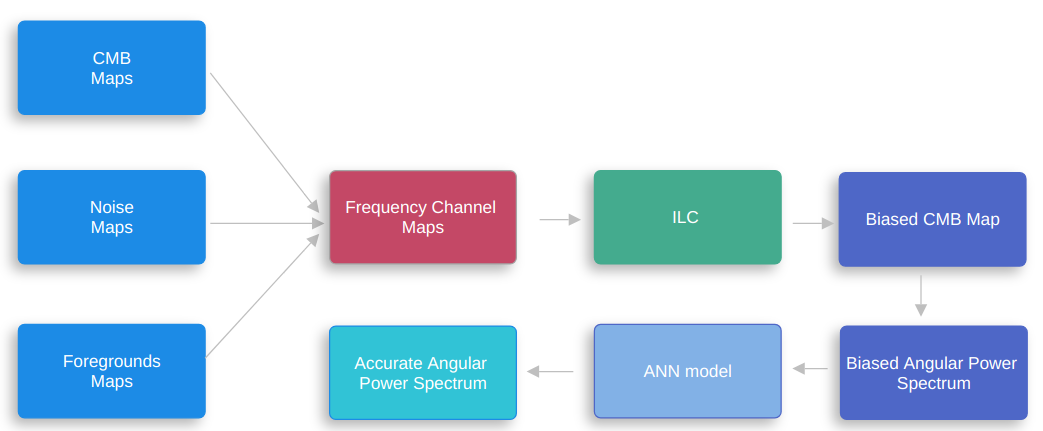}
    \caption{In the block diagram above, we show the workflow of our method. The input maps—comprising CMB, noise, and foreground maps—are combined to generate frequency channel maps. These maps are processed through the Internal Linear Combination (ILC) method, resulting in a biased CMB map and a biased angular power spectrum. An Artificial Neural Network (ANN) model is then applied to correct these biases, producing an accurate angular power spectrum.}
    \label{workflow}
\end{figure*}

%% Methods
\section{Formalism}
\label{sec:formalism}

The proposed methodology, as illustrated in Figure \ref{workflow}, aims to develop a hybrid approach for obtaining an accurate CMB B-mode angular power spectrum over large angular scales ($2 \leq \ell \leq 47$). This method involves combining input maps of the polarized Cosmic Microwave Background (CMB), noise, and foregrounds to generate frequency channel maps that serve as the basis for analysis. An Internal Linear Combination (ILC) approach is then applied to these maps to extract a foreground-minimized CMB B-mode map, which contains residual biases. From the ILC maps, we obtain the angular power spectrum, which again is biased. To minimize this complex bias, we train an Artificial Neural Network (ANN) model. 

In the following subsections, we provide a detailed overview of our methodology. We begin with the data model, which describes how observations of the CMB are represented using Stokes parameters and transformed into the B-mode maps. Next, we discuss the Internal Linear Combination (ILC) method, with a focus on the `global' pixel space ILC, which is used in the current investigation to extract the CMB signal. Following this, we discuss the biases in the ILC approach and the challenges they present in achieving accurate signal recovery. We then introduce the ANN framework, which is designed to mitigate these biases and enhance the precision of the reconstructed CMB B-mode angular power spectrum. Finally, we describe the simulation framework employed in our study, including the generation of CMB, noise, and foreground components, which are critical for training and validating our proposed hybrid methodology.

\subsection{Data Model}
Linear polarization of the CMB are observed as the Stokes parameters \(Q\) and \(U\), which are coordinate-dependent quantities. When the polarization is expressed as a complex number \(P = Q \pm iU\), a rotation of the coordinate axes by an angle \(\alpha\) results in a transformation \(P \rightarrow P e^{\pm 2i\alpha}\), implying that \(P\) is a spin-2 field \cite{1997NewA....2..323H,2016ARA&A..54..227K}. This spin-2 field can be expanded on the surface of a sphere using spin-2 spherical harmonics \( _{\pm 2}Y_{\ell m}(k) \):

\begin{equation}
Q(k) \pm iU(k) = \sum_{\ell m} a_{\pm 2,\ell m} ~_{\pm 2}Y_{\ell m}(k),
\end{equation}

where \(k\) is the pixel index and \(a_{\pm 2, \ell m}\) are spin-2 spherical harmonic coefficients.

We define the spin-0 B-mode map, which is coordinate-independent, as:

\begin{equation}
B(k) = \sum_{\ell m} a_{\ell m}^B Y_{\ell m}(k),
\end{equation}

where the B-mode spherical harmonic coefficients are given by

\begin{equation}
a_{\ell m}^B = \frac{1}{2} (a_{+2, \ell m} - a_{-2, \ell m}).
\end{equation}

Working in the spin-0 harmonic basis, the full-sky \(Q\) and \(U\) maps for each frequency channel are transformed to the corresponding B-mode maps. Given the CMB B-mode signal \(S\) observed across \(n\) frequency channels, the observed map for the \(i\)-th frequency channel \(X_i\) is modeled as:
\begin{equation}
\mathbf{X}_i = \mathbf{S} + \mathbf{F}_i + \mathbf{N}_i ,  
\label{eq:X_i}
\end{equation}
where $\mathbf{S}$ is the CMB signal, $\mathbf{F}_i$ is the foreground contamination, and $\mathbf{N}_i$ is the noise at each frequency channel. Each of these quantities is a vector of size \(N_{pix}\), representing a HEALPix \cite{2005ApJ...622..759G} map, where \(N_{pix} = 12N_{side}^2\), and \(N_{side}\) is the pixel resolution parameter. The observed data set is thus \( \mathbf{D} = \{\mathbf{X}_1, \mathbf{X}_2, ..., \mathbf{X}_n\} \).

%=======================
\subsection{Internal Linear Combination}
\label{ilc}

The ILC method is a widely used technique to extract the CMB signal by minimizing the contamination from foregrounds. The basic idea is to form a linear combination of observed frequency maps that reduces the variance, thus suppressing foregrounds and noise while preserving the CMB signal, which is assumed to be frequency-independent.

The ILC method can be implemented in different spaces, such as pixel domains \cite{2003ApJS..148...97B,2004ApJ...612..633E}, harmonic space \cite{2003PhRvD..68l3523T}, and needlet space \cite{2009A&A...493..835D}. The pixel-based ILC method forms linear combinations of the frequency bands. The weights \(w_i\) are either pixel dependent or are independent of pixels for each frequency channel. In the later case it is known as the ``global ILC" (see e.g. \cite{2007astro.ph..2198D}). In harmonic space, the ILC is applied in the spherical harmonic domain. The frequency channel maps are decomposed into spherical harmonics, and the linear combination of frequency maps is performed in multipole space. In harmonic domain the weights are function of multipole. This allows for operations to be performed on the harmonic coefficients, reducing computational costs and providing a natural scale-by-scale foreground removal. Needlets provide a wavelet basis on the sphere that offers localization in both pixel and harmonic space. The ILC in needlet space takes advantage of this property to better handle foregrounds and noise at different angular scales by performing the ILC independently in different needlet bands, which provides better control over the localization of foreground minimization.

In this work we use, the global ILC to form a weighted sum of the observed frequency maps to minimize the variance of the combined map. Suppose we have observations at $n$ different frequencies $\mathbf{D}$, with observations at each frequency given by Equation \ref{eq:X_i}, then the combined map $\mathbf{S}$ can be written as:
\begin{equation}
\hat{\mathbf{S}} = \sum_{i=1}^{n} w_i \mathbf{X}_i    
\end{equation}
where $w_i$ are the weights applied to each frequency map and $\hat{\mathbf{S}}$ is the estimated CMB signal. The ILC aims to find these weights by minimizing the total variance of the combined map while preserving the CMB signal.

The total variance of the combined map is given by:
\begin{equation}
\text{Var}(\mathbf{S}) = \left\langle \left(\sum_{i=1}^{n} w_i \mathbf{X}_i \right)^2 \right\rangle
\end{equation}
The ILC weights are determined by minimizing this variance, subject to the constraint that the sum of the weights preserves the CMB signal:
\begin{equation}
\sum_{i=1}^{n} w_i = 1    
\end{equation}
This optimization problem can be solved using Lagrange multipliers. Let the covariance matrix $\mathbf{C}$ between the frequency maps $\mathbf{X}_i$ and $\mathbf{X}_j$ be defined as:
\begin{equation}
C_{ij} = \langle \mathbf{X}_i \mathbf{X}_j \rangle    
\end{equation}
The weights that minimize the variance are given by:
\begin{equation}
w_i = \frac{\sum_{j=1}^{n} C^{-1}_{ij}}{\sum_{i=1}^{n} \sum_{j=1}^{n} C^{-1}_{ij}}    
\end{equation}
The ILC weights are applied to the observed maps to obtain the foreground-minimized CMB map $\hat{\mathbf{S}}$.
By minimizing the variance of the combined map, the contributions from foregrounds and noise are reduced, leading to a map dominated by the CMB signal. The method works under the assumption that the CMB signal is frequency-independent, though residual noise and foreground biases may persist for the reasons discussed in the next section.

%================================

 \subsection{Bias in ILC}
 \label{bias}
%%The ILC bias arises due to incomplete separation of the CMB signal from noise and foregrounds, during the variance minimization process. One of the crucial assumption in ILC method is that the CMB is uncorrelated with foreground or noise. The ILC method, designed to reduce the total variance in the observed data, suppresses parts of the CMB signal because of correlations with noise or foregrounds. These empirical correlations exist due to chance correlations between CMB and foreground components on a particular realization of CMB sky [cite]. This bias depends on the relative strength of the signal compared to the noise and foregrounds. When the CMB signal is weak, as in the case of B-modes, the observed data is dominated by noise and foregrounds, making the separation more difficult. As a result, the ILC method can mistakenly treat the weak signal as part of the contaminants, leading to higher bias in the recovered map. In such cases, inaccuracies in the empirical covariance matrix estimation further amplify the bias, making the reconstruction process less reliable. Thus, the bias increases as the signal becomes weaker relative to the contaminants, directly affecting the precision of the CMB signal extraction.%%

When dealing with a weak CMB B-mode signal in the presence of strong foregrounds and noise, the ILC method can produce biased CMB map \cite{2009A&A...493..835D,Saha:2007gf}.  The sources of bias in the ILC map output under these conditions can be identified as follows:

\begin{enumerate}
    \item \textbf{Foreground Residuals}: Since the foregrounds are strong and correlated across frequency channels, the ILC method may not fully remove them. This leads to residual foreground contamination in the reconstructed CMB map, as the minimization process is affected by the dominant foreground component, introducing bias.
    
    \item \textbf{Chance Correlation}: Empirical correlations can arise due to chance alignments between the CMB and foreground components and noise in a particular realization of the CMB sky \cite{}. This violates the assumption that the CMB is uncorrelated with foregrounds or noise, leading to additional bias in the ILC output.
\end{enumerate}

\subsection{ Artificial Neural Network}

Artificial Neural Networks (ANNs) are known to approximate complex functions \( f: \mathbb{R}^n \to \mathbb{R}^m \) by learning directly from data \cite{HORNIK1991251,1999AcNum...8..143P}. An artificial neuron, the fundamental unit of an ANN, is a mathematical function that mimics the behavior of a biological neuron \cite{Bishop}. While a single neuron has limited capability, combining many neurons in a network enables it to learn and identify complex patterns within data. The model parameters—weights and biases—are adjusted to minimize a defined loss function \( \mathcal{L} \), allowing the network to capture and generalize patterns through input-output mappings in supervised learning tasks.

In this article, we utilize a supervised, fully connected deep artificial neural network (ANN). Supervised learning trains the model with labeled data, helping it learn accurate mappings by reducing the error between predicted and actual values. Deep learning, characterized by networks with multiple hidden layers, allows the model to capture hierarchical features. In a fully connected (dense) network, each neuron in one layer connects to every neuron in the next layer, enhancing the network’s ability to learn intricate relationships within the data.

\subsubsection*{Neural Network Architecture}

\begin{figure}[ht]
    \centering
    \includegraphics[width=.5\textwidth]{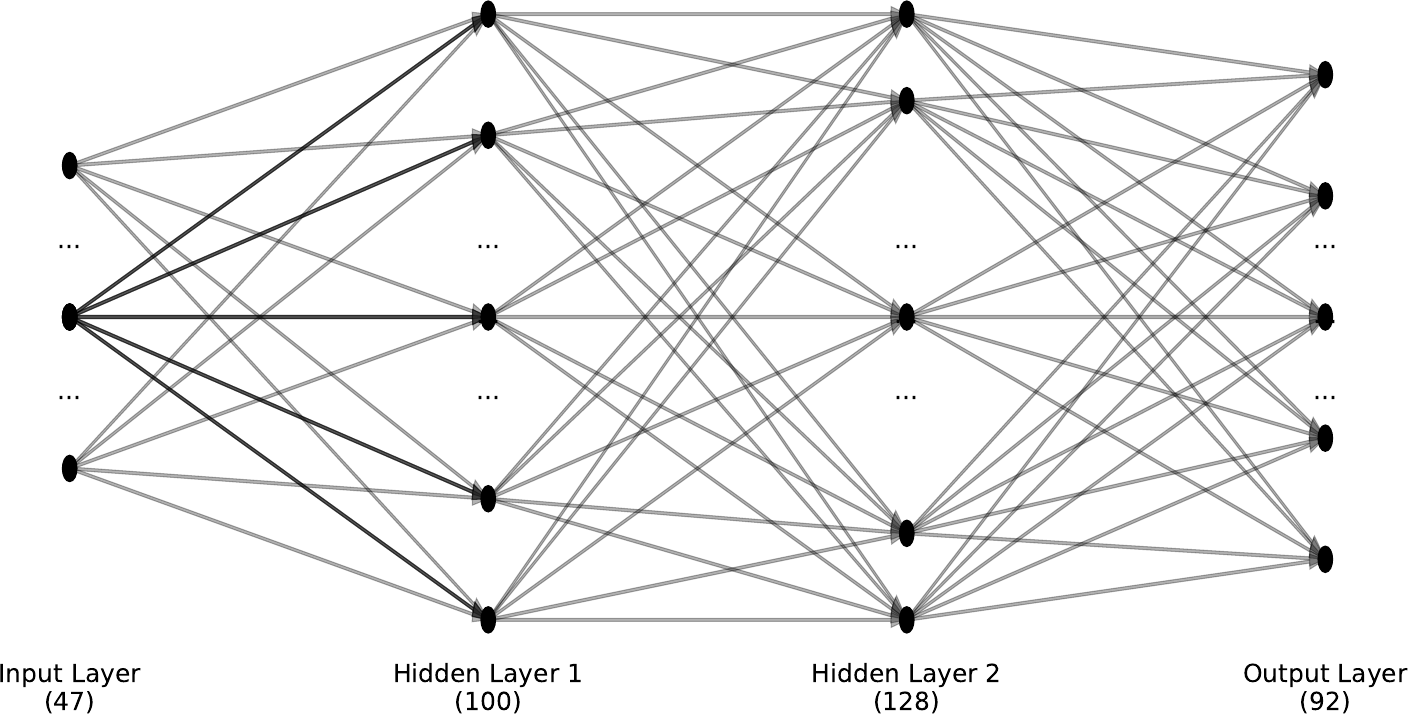}
    \caption{Architecture of the neural network used in our study. The network consists of an input layer with 47 nodes, followed by two hidden layers with 100 and 128 neurons, respectively, and an output layer with 92 nodes. Each dot represents a neuron, while the connecting lines illustrate the fully connected nature of the network, where each neuron in one layer is linked to every neuron in the next layer. This configuration enables the model to capture complex features for accurate predictions in CMB B-mode power spectrum reconstruction.}
    \label{fig:nn}
\end{figure}

The neural network architecture is organized in layers, with each layer transforming the input through weighted sums and activation functions. In our study, the architecture, shown in Figure \ref{fig:nn}, comprises an input layer, an output layer, and two hidden layers in between. Each circular unit represents a 'neuron,' and the connecting lines represent the parameters (weights and biases), which are adjusted during training to optimize performance.

An input \( \mathbf{x} \) applied to a neuron produces an output \( \mathbf{y} \) through the transformation:
\begin{equation}
\mathbf{y} = \sigma(\mathbf{w} \cdot \mathbf{x} + \mathbf{b}),
\end{equation}
where \( \mathbf{w} \) and \( \mathbf{b} \) denote the weight and bias, respectively. The activation function \( \sigma \) introduces non-linearity, allowing the network to learn complex patterns.

The input layer consists of 46 features, followed by two fully connected hidden layers with 50 and 128 neurons, respectively. Finally, the \textbf{output layer} has 92 neurons, generating \( \hat{\mathbf{y}} \) as an approximation of the target \( \mathbf{y} \), with additional outputs to represent predicted variance or uncertainty.

\subsubsection*{Heteroscedastic Loss function}

Precise uncertainty estimation is essential in scientific studies, as it provides insights into the reliability of model predictions. Uncertainty can generally be categorized into \textbf{aleatoric} and \textbf{epistemic} types \cite{2017arXiv170304977K}. Aleatoric uncertainty arises from inherent noise in the data and is therefore \textbf{irreducible}—additional data cannot eliminate it. Epistemic uncertainty, by contrast, stems from limited knowledge about the model itself, particularly its parameters, and is \textbf{reducible} with more data or a better model (Kendall \& Gal 2017). To quantify aleatoric uncertainty in our model, we employ a heteroscedastic loss function \cite{2017arXiv170304977K}, which allows the model to learn and predict varying levels of uncertainty across different data points:

\begin{equation}
 \mathcal{L} = \frac{1}{2n_{out}} \sum_{k=1}^{n_{out}} \left[ \exp(-s_{k}) (y_{k} - \hat{y}_{k})^2 + s_{k} \right]    
\end{equation}
where \( s_{k} = \log(\sigma_{k}^{2}) \) and \( \sigma_{k}^{2} \) represents the aleatoric uncertainty associated with the predicted value \( \hat{y}_{k} \).

\subsubsection*{Training}

For training the model, the weights and biases in the network are first randomly assigned; we use He uniform initialization in our case, which helps maintain stable variance in activations across layers, reducing the risk of vanishing or exploding gradients and improving convergence speed in deep networks.

Suppose we have labeled sample with input set \( \mathbf{X} \) and corresponding target \( \mathbf{Y} \). 
Forward propagation computes the output \( \hat{\mathbf{Y}}^{k} \) by sequentially passing the input \( \mathbf{X}^{k} \) through each layer for a given sample $k$. For each layer \( l = 1, \dots, L \), the transformations are applied as follows:

\begin{enumerate}
    \item Compute the weighted sum:
    \begin{equation}
    \mathbf{X}^{(k)}_{l} = \mathbf{W}^{(k)}_{l} \mathbf{X}^{(k)}_{l-1} + \mathbf{B}^{(k)}_{l}    
    \end{equation}
    
    \item Apply the activation function:
    \begin{equation}
    \mathbf{X}^{(k)}_{l} = \sigma(\mathbf{X}^{(k)}_{l})        
    \end{equation}

\end{enumerate}

The output layer provides \( \hat{\mathbf{Y}}^{k} = \mathbf{X}^{(k)}_{L} \), which is the network’s prediction for $k^{th}$ sample in input \( \mathbf{X} \).

Once we pass all the training set through the network, we obtain the cost function: 

\begin{equation}
       \mathcal{J} = \frac{1}{m} \sum_{k=1}^{m} \mathcal{L}_{k} 
\end{equation}
where $\mathcal{L}_{k}$ is the loss for $k^{th}$ sample and m is the total number of training samples.

The network’s weights and biases are updated to minimize the cost function in a process called \( Error ~Back ~Propagation \) \cite{Hecht-Nielsen}. It involves computing the gradient of the cost function \( \mathcal{J} \) with respect to weights and biases by using the chain rule:

\begin{equation}
\delta{w_{ij}} = \frac{\partial\mathcal{J}}{\partial w_{ij}}    
\end{equation}

\begin{equation}
\delta{b_{ij}} = \frac{\partial\mathcal{J}}{\partial b_{ij}}.    
\end{equation}

At the end of every iteration an \( Optimization ~Algorithm \) updates  $w_{ij} \rightarrow w_{ij} + \alpha\delta{w}_{ij} $ and $b_{ij} \rightarrow b_{ij} + \alpha\delta{b}$. In our case we use \( adaptive ~moment ~estimation \) algorithm \cite{2014arXiv1412.6980K}. The $\alpha$ is called the learning rate and is a parameter associated with the optimizer.  

Updating neural network weights using gradients from the entire training set in each iteration can slow learning and increase the chance of getting trapped in local minima. To counter this, mini-batches of size \( m_b \) (where \( m_b < m \)) are used with algorithms like Mini-batch Stochastic Gradient Descent (MSGD) and ADAM \cite{2014arXiv1412.6980K}, which calculate gradients from subsets of data, enabling the network escape local minima. An \textbf{epoch} represents one full pass through the dataset, and shuffling the data before each epoch improves generalization.

In our study, 80\% of the dataset is used for training, and 10\% each is reserved for validation and testing. To handle the wide range of power spectrum values, input features are logarithmically scaled and normalized. We use the \texttt{Adam} optimizer with a 0.00065 learning rate, alongside \textbf{early stopping} and \textbf{model checkpointing} to prevent overfitting and retain the best model. Training halts if validation loss doesn’t improve for 10 epochs, with up to 1500 epochs and a batch size of 4096.

\subsection{Simulations}
\label{subsec:sim}
 This section details the simulation procedures used to generate various polarized component contributing to the microwave sky - the CMB B-mode map, noise realizations, and foregrounds. We use these simulations to generate frequency channel maps at proposed \textsc{ECHO}\footnote{\url{https://cmb-bharat.in/}} mission.  We subsequently use the simulated frequency channel maps as input to ILC pipeline to obtain foreground minimized maps. We use the angular power spectrum from the ILC maps next as an input to train an ANN.  

\begin{table}
\hspace{-1cm}
   \begin{tabular}{ p{2.cm}  p{2.cm}   p{3.2cm}  }
   \hline
Frequency & $\theta_{\mathrm{FWHM}}$ & Noise $\Delta_P$  \\
(GHz)     & (arcmin)                 & ($\mu$K-arcmin)   \\
   \hline
   28&39.9&16.5\\
   35&31.9&13.3\\
   45&24.8&11.9\\
   65&17.1&8.9\\
   75&14.91&5.1\\
   95&11.7&4.6\\
   115&9.72&3.1\\
   130&8.59&3.1\\
   145&7.70&2.4\\
   165&6.77&2.5\\
   190&5.88&2.8\\
   220&5.08&3.3\\
   275&4.06&6.3\\
   340&3.28&11.4\\
   390&2.86&21.9\\
   450&2.48&43.4\\
   520&2.14&102.0\\
   600&1.86&288.0\\
   700&1.59&1122.0\\
   850&1.31&9550.0\\
 \hline
   \end{tabular}
\caption{Instrumental specification for \ECHO\ as in the \cmbbh\ proposal. We show the frequency band centers, FWHM beam apertures, polarization white-noise levels.}
\label{tab:noise}
\end{table}

 \begin{figure}[ht]
    \centering
    \includegraphics[width=0.5\textwidth]{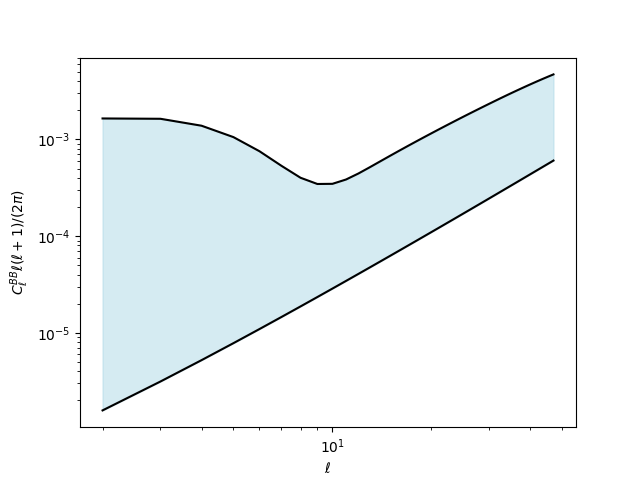}
    \caption{In figure we show the range of theoretical CMB B-mode angular power spectra for the chosen tensor-to-scalar ratio \( r \) and reionization optical depth \( \tau \) parameters used in this work. The shaded region represents the range within which the theoretical angular power spectrum lies, based on the specified \( r \) and \( \tau \) values.
}
    \label{fig:param}
\end{figure}

\begin{figure}[ht]
    \centering
    \includegraphics[width=0.5\textwidth]{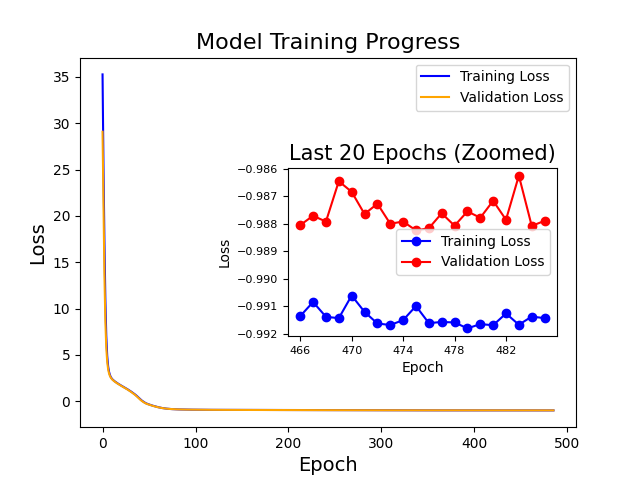}
    \caption{Model training progress showing the loss as a function of epochs. The main plot displays both the training loss (blue) and validation loss (orange) over 500 epochs, demonstrating the model’s convergence. The inset plot zooms in on the last 20 epochs, highlighting the stability and small fluctuations in the training and validation losses as the model approaches optimal performance.}
    \label{fig:loss_vs_epoch}
\end{figure}

\begin{figure}[ht]
    \centering
    \includegraphics[width=0.5\textwidth]{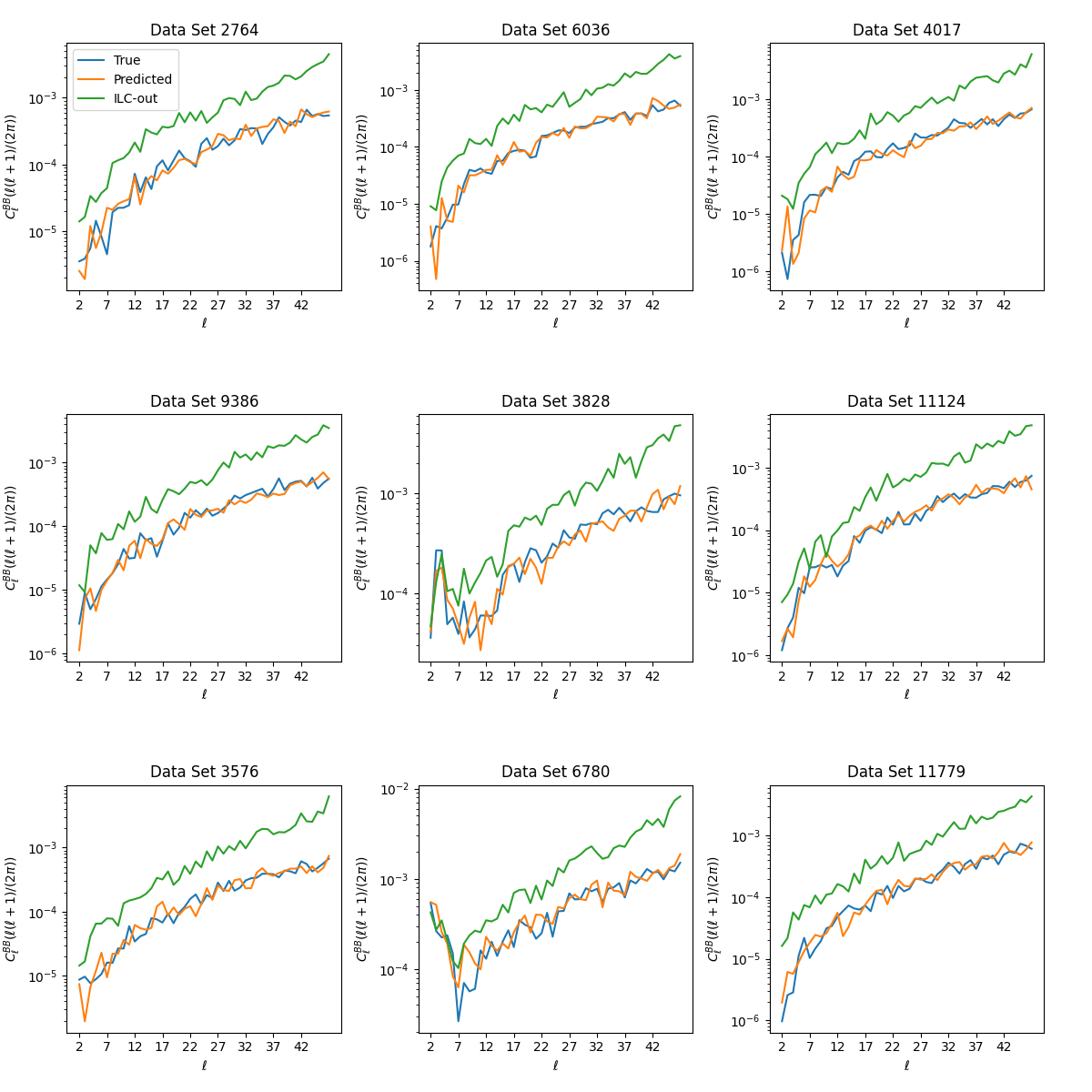}
    \caption{Comparison of the true, predicted, and ILC-derived CMB B-mode angular power spectra \( C_{\ell}^{BB}\) across randomly selected datasets from the test set. Each panel represents a unique dataset, with the true spectrum shown in blue, the predicted spectrum in orange, and the ILC output in green. The predicted spectra closely follow the true spectra across various multipole moments \( \ell \), while the ILC output exhibits higher deviations. This demonstrates the model's ability to generalize effectively on unseen data and provides more accurate reconstructions of the true CMB B-mode power spectrum compared to the ILC method.
}
    \label{fig:random_plots}
\end{figure}

\begin{figure}[ht]
    \centering
    \includegraphics[width=0.5\textwidth]{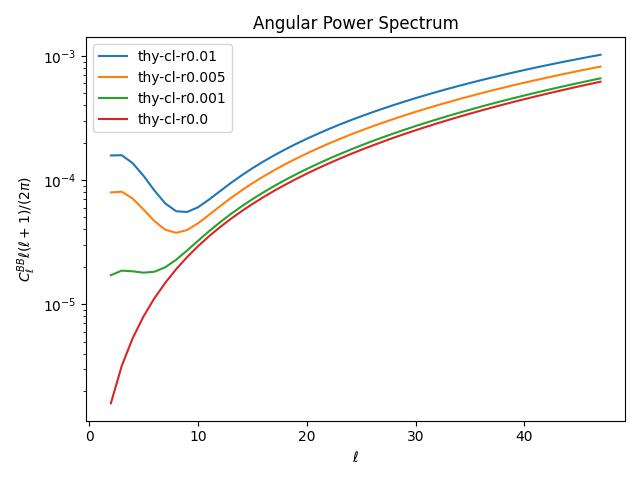}
    \caption{Theoretical CMB B-mode angular power spectrum \( C_{\ell}^{BB}\) for different tensor-to-scalar ratio (\( r \)) values: \( r = 0.01 \), \( r = 0.005 \), \( r = 0.001 \), and \( r = 0.0 \). Each curve represents the theoretical angular power spectrum for a specific \( r \) value, illustrating how the B-mode signal amplitude varies with \( r \) across multipole moments \( \ell \). These four theoretical spectra are used to quantify the performance of our method.
}
    \label{fig:test_case}
\end{figure}

\begin{figure}[ht]
    \centering
    \includegraphics[width=0.5\textwidth]{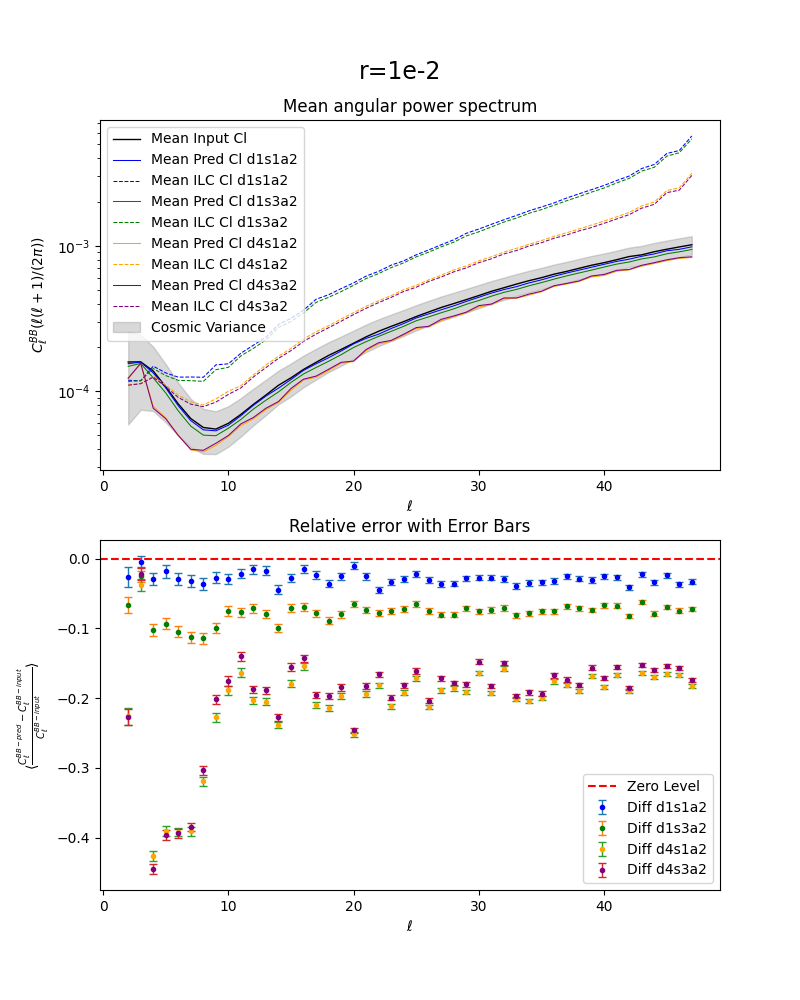}
    \caption{Comparison of the mean CMB B-mode angular power spectrum \( C_{\ell}^{BB} \) for \( r = 10^{-2} \) across different foreground models. The top panel shows the mean angular power spectra for the input (black), predicted (solid lines), and ILC (dashed lines) across various foreground scenarios: \( d1s1a2 \), \( d1s3a2 \), \( d4s1a2 \), and \( d4s3a2 \). The shaded gray region represents the \( 1\sigma \) cosmic variance, serving as a reference for acceptable uncertainties. The bottom panel displays the relative error between the predicted and true spectra for each foreground model, with error bars indicating variability. The red dashed line at zero represents the ideal error-free level, and deviations indicate the impact of foreground complexity on model performance. The predicted spectra show smaller deviations compared to ILC, demonstrating improved accuracy and robustness across different foreground conditions.
}
    \label{fig:clr10-2}
\end{figure}

\begin{figure}[ht]
    \centering
    \includegraphics[width=0.5\textwidth]{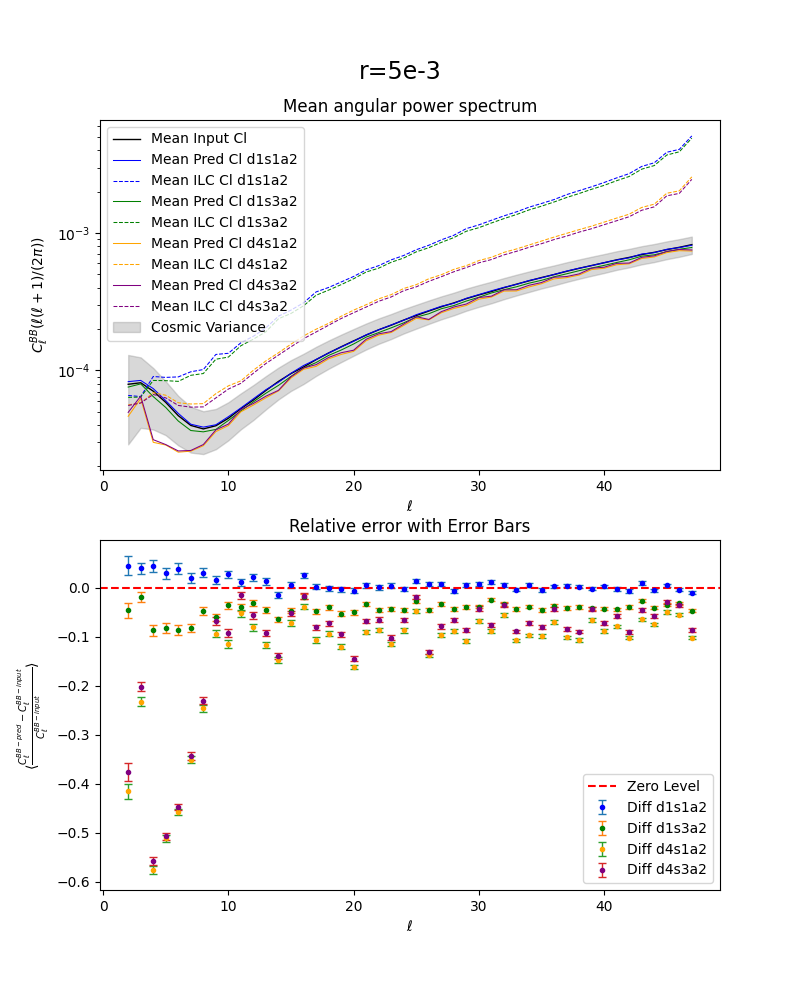}
    \caption{Mean angular power spectrum and relative error comparison for \( r = 5 \times 10^{-3} \). The predicted spectra remain close to the true spectrum across the different foreground models, while the ILC output continues to show larger deviations, especially at higher multipoles. The relative error plot reveals that the predicted model achieves better consistency than the ILC, with error bars that show minor increases in deviation compared to the \( r = 10^{-2} \) case, particularly at lower \( \ell \) values.}
    \label{fig:clr510-3}
\end{figure}

\begin{figure}[ht]
    \centering
    \includegraphics[width=0.5\textwidth]{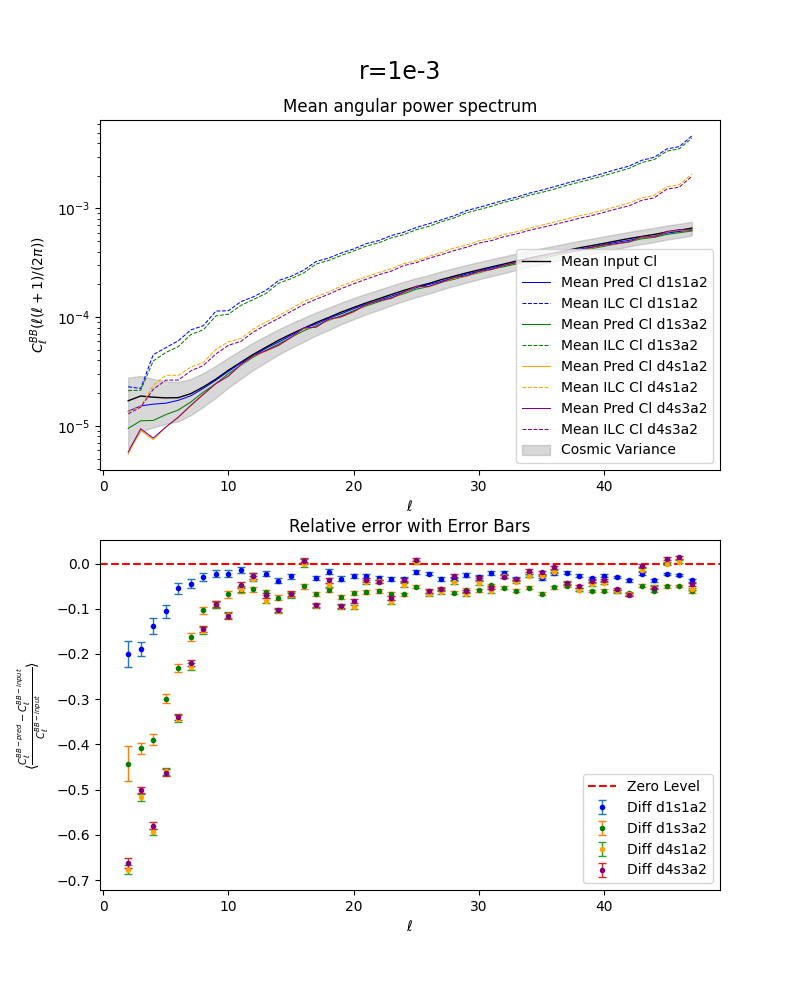}
    \caption{Mean angular power spectrum and relative error comparison for \( r = 10^{-3} \). The predicted spectra retain close alignment with the true spectra, even as the tensor-to-scalar ratio decreases, while the ILC output still exhibits noticeable deviations. In the relative error plot, the predicted model demonstrates robustness, though deviations increase slightly at low multipoles (\( \ell < 10 \)), which is expected given the weaker signal. Error bars indicate that the predicted model still outperforms the ILC, maintaining relative accuracy across all foreground combinations.
}
    \label{fig:clr10-3}
\end{figure}

\begin{figure}[ht]
    \centering
    \includegraphics[width=0.5\textwidth]{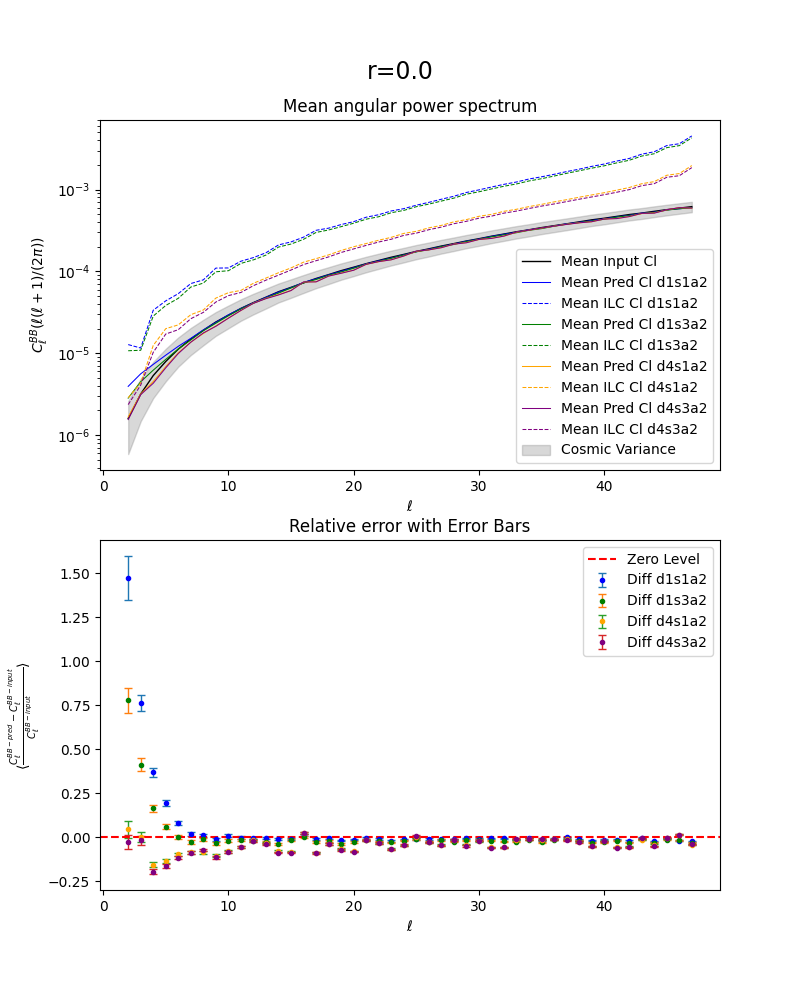}
    \caption{Mean angular power spectrum and relative error comparison for \( r = 0.0 \), representing a null B-mode signal scenario. The predicted spectra remain aligned with the true spectrum, while the ILC output continues to diverge. The relative error plot shows minimal deviations for the predicted model, even with an absent primordial signal, confirming its effectiveness for null tests. Error bars at low multipoles reveal minor variability, but the model still maintains a higher accuracy than the ILC, demonstrating robustness across foreground scenarios in the absence of a tensor component.}
    \label{fig:clr0}
\end{figure}

\begin{figure}[ht]
    \centering
    \includegraphics[width=0.5\textwidth]{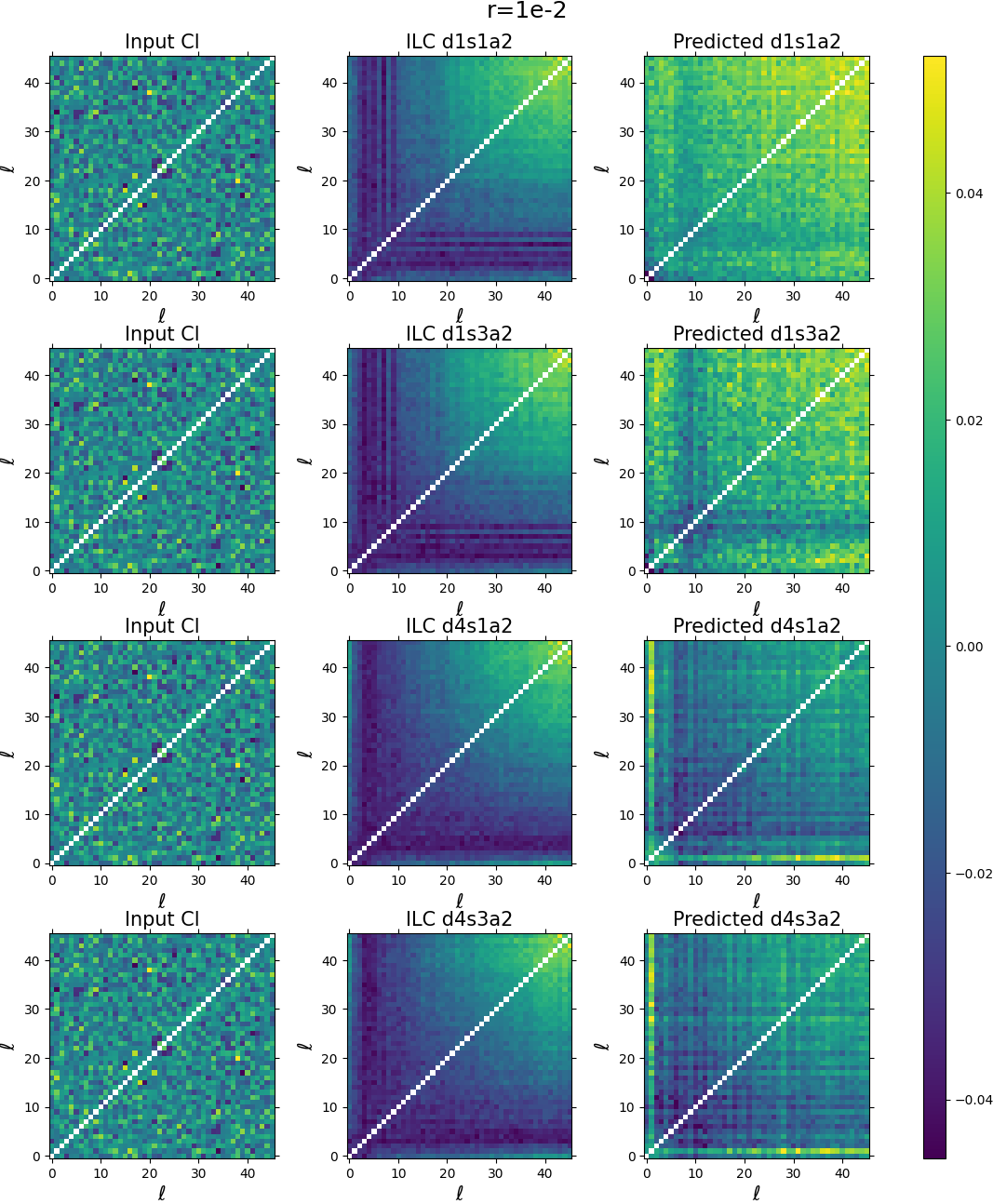}
    \caption{Correlation matrices for \( r = 10^{-2} \) across different foreground models (\( d1s1a2 \), \( d1s3a2 \), \( d4s1a2 \), and \( d4s3a2 \)). Each row represents a specific foreground, with columns showing the input \( C_\ell \), ILC output, and predicted output from the model. The predicted matrices closely match the input structure, while the ILC results display stronger off-diagonal correlations, particularly with more complex foregrounds. This highlights the ANN based method is effective in preserving the true correlation structure w.r.t the ILC method.
}
    \label{fig:correlr10-2}
\end{figure}

\begin{figure}[ht]
    \centering
    \includegraphics[width=0.5\textwidth]{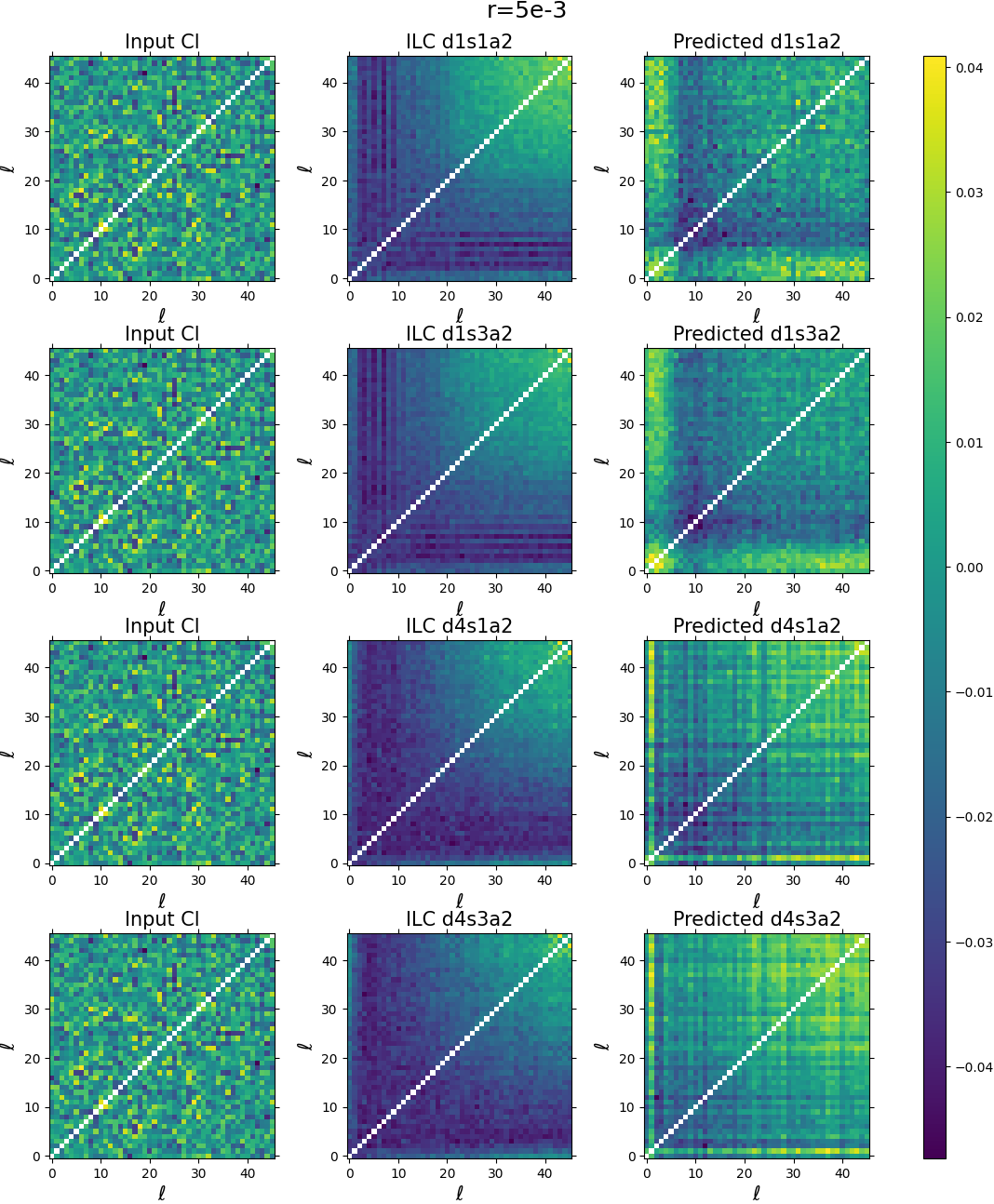}
    \caption{For \( r = 5 \times 10^{-3} \), the correlation matrices are shown for the same foreground scenarios as before. The predicted output maintains close alignment with the input structure, particularly in foreground cases \( d1s1a2 \) and \( d1s3a2 \). In contrast, the ILC output introduces notable off-diagonal correlations, which are more evident as foreground complexity increases (e.g., \( d4s1a2 \) and \( d4s3a2 \)). This consistency underscores the ANN based method's robustness as the B-mode signal amplitude decreases.
}
        \label{fig:correlr510-3}
\end{figure}

\begin{figure}[ht]
    \centering
    \includegraphics[width=0.5\textwidth]{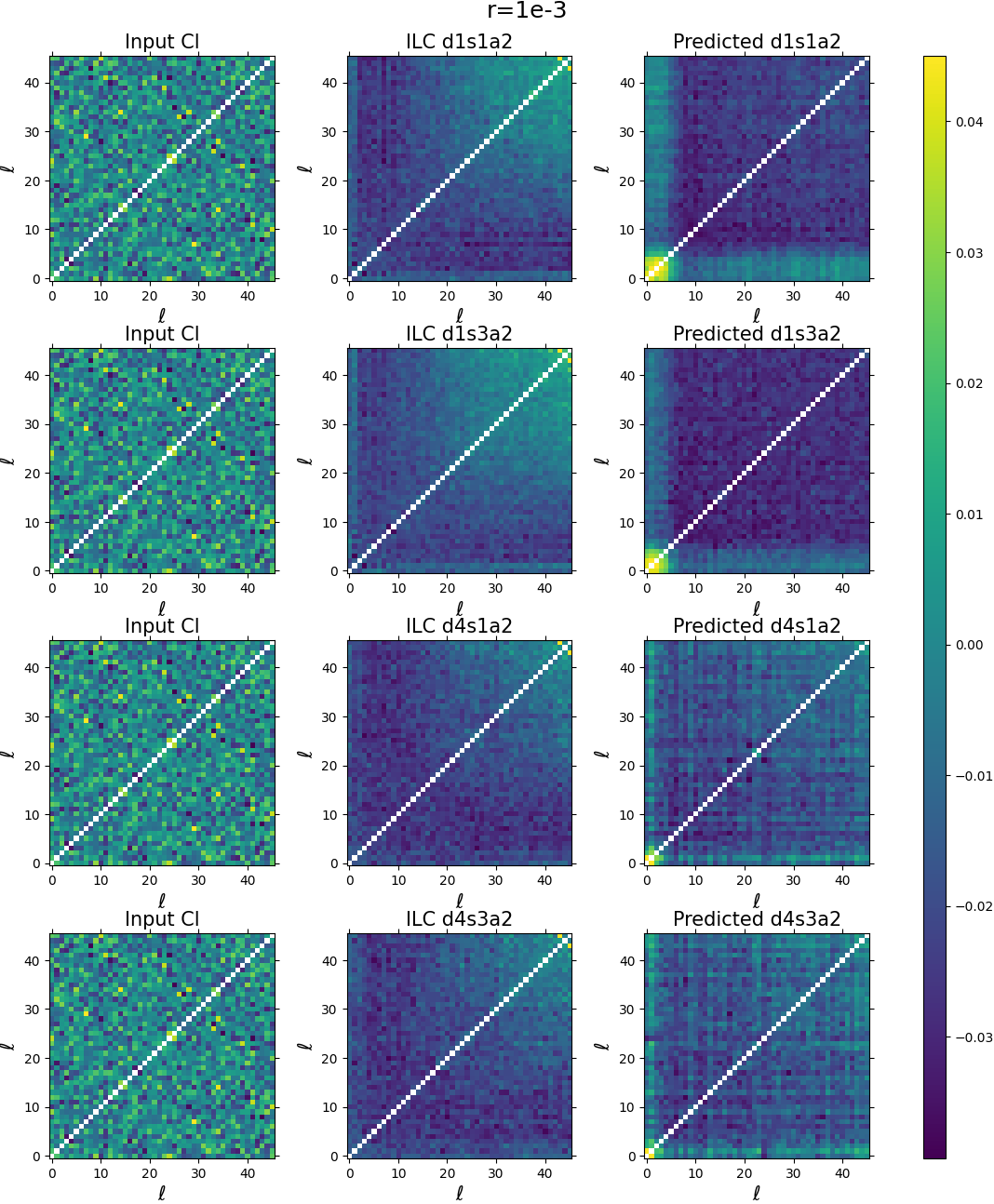}
    \caption{Correlation matrices for \( r = 10^{-3} \), displaying input \( C_\ell \), ILC output, and model predictions for each foreground scenario. The predicted matrices continue to closely follow the input structure, whereas the ILC matrices show amplified off-diagonal correlations in cases with complex foregrounds. The ANN method demonstrates stability in retaining accurate cross-correlation patterns even at this lower \( r \) value, where the B-mode signal is weaker.}
        \label{fig:correlr10-3}
\end{figure}

\begin{figure}[ht]
    \centering
    \includegraphics[width=.5\textwidth]{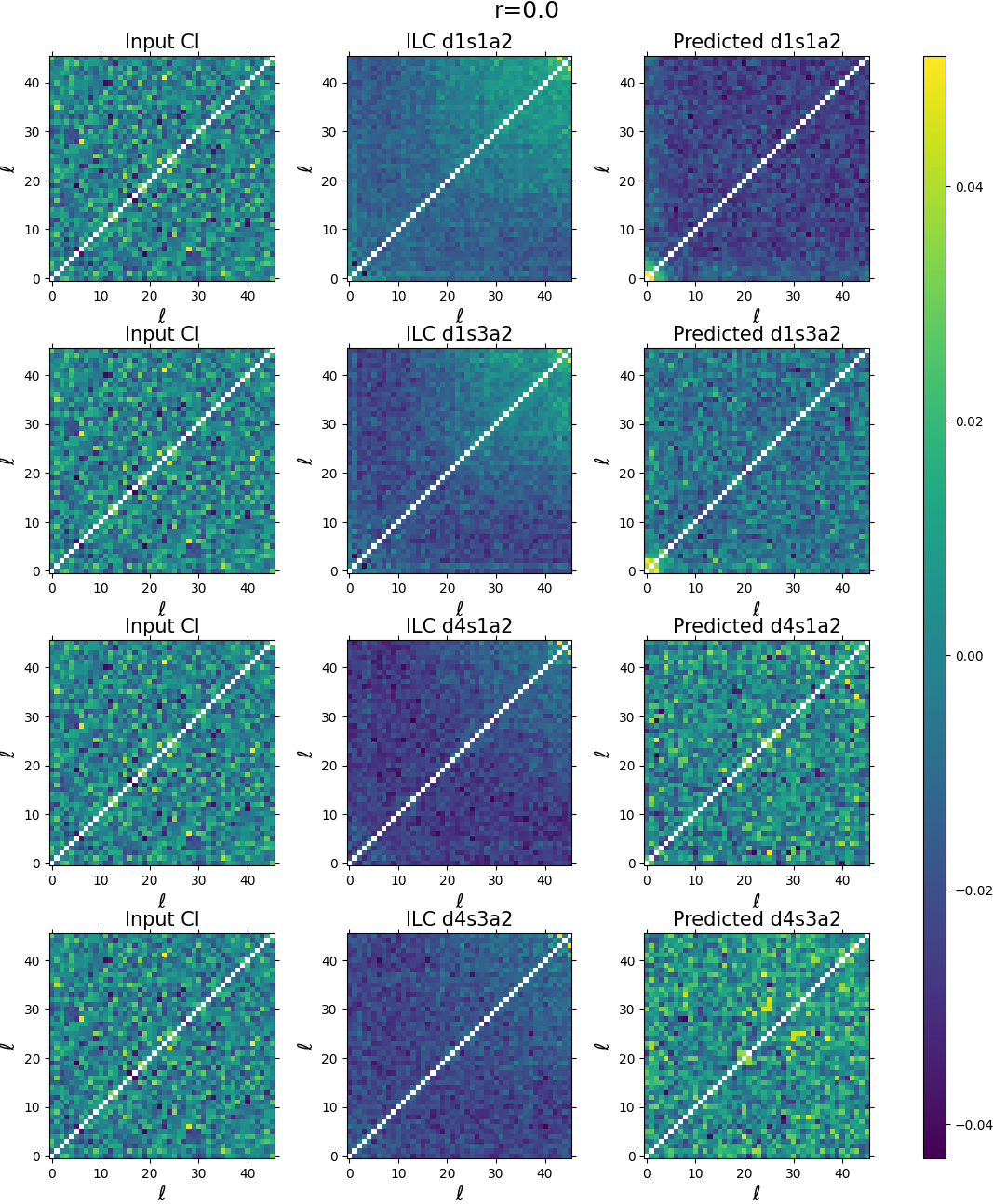}
    \caption{Correlation matrices for \( r = 0.0 \) (null signal case) across the four foreground models. The predicted matrices still closely resemble the input structure, effectively capturing the true correlation structure despite the absence of a primordial B-mode signal. Notably, the network was never trained on the \( r = 0.0 \) case, yet it successfully generalizes to this null scenario. The ILC output, however, shows prominent off-diagonal correlations across all foreground scenarios, indicating residual contamination. The ANN method’s performance in this null case further emphasizes its reliability in minimizing artificial correlations and validating null tests.}
    \label{fig:correlr0}
\end{figure}

\subsubsection{CMB Simulations}
To generate the CMB B-mode maps and their corresponding angular power spectra, we sample the tensor-to-scalar ratio \(r\) in multiple ranges: \([0.1, 0.01]\), \([0.01, 0.001]\), \([0.001, 0.0001]\), \([0.0001, 0.00001]\), and \([0.00001, 0.000001]\). The values within each range is sampled uniformly. A total of 2000 samples are generated from each range, resulting in a comprehensive set of \(r\) values for training and validation.

Using sampled \( r \) values and a fixed reionization optical depth \( \tau = 0.0561 \), we employ CAMB \cite{2000ApJ...538..473L} to compute the theoretical angular power spectra \( C_{\ell}^{BB} \) for each \( r \). Figure \ref{fig:param} shows the range of CMB B-mode theoretical angular power spectra for the chosen \( r \) and \( \tau \) parameters in this work. The resulting power spectra are stored and used to generate the CMB B-mode maps using the \texttt{synfast} facility in \texttt{HEALPix}. These B-mode maps are subsequently smoothed with a \( 9^\circ \) beam and pixel window function. For each theoretical \( C_{\ell}^{BB} \), we generate 15 map simulations and corresponding angular power spectra. Overall, the CMB simulations yield 150,000 maps and angular power spectra.

%%%%%%%%%%%%%%%%%%%%%%%%%%%%%%%%%%%%%%%%%%%%%%%%%%%%%%%%%%%%
% Section 3.3: Detector Noise Simulations
%%%%%%%%%%%%%%%%%%%%%%%%%%%%%%%%%%%%%%%%%%%%%%%%%%%%%%%%%%%%
\subsubsection{Detector Noise Simulations}
To simulate the detector noise, we generated Gaussian, isotropic, and pixel-uncorrelated realizations of random noise for the Stokes parameters $Q$ and $U$ across the twenty \ECHO frequency bands. The detailed detector specifications for each frequency band are provided in Table \ref{tab:noise} . We assumed that the $Q$ and $U$ noise maps are pixel-uncorrelated, such that:

\begin{equation}
    \langle Q_{i}(p) U_{i}(p') \rangle = 0,
\end{equation}

where $Q_i$ and $U_i$ represent the Stokes parameters for the frequency channel $i$, and $p$ and $p'$ are pixel indices. The noise variances for $Q$ and $U$, denoted as $\sigma^2_{Q_i}$ and $\sigma^2_{U_i}$ respectively, are assumed to be identical and can be expressed as:

\begin{equation}
    \sigma^2_{Q_i} = \sigma^2_{U_i} = \Delta Q^2_i \zeta^2 \Delta \Omega,
\end{equation}

where $\Delta Q_i$ is the root mean square (rms) noise value in arcminutes for the $Q_i$ map, $\zeta$ is a conversion factor from arcminutes to radians, and $\Delta \Omega$ is the solid angle of a pixel at $N_{\text{side}} = 16$. Both $Q$ and $U$ noise maps were adjusted to the same beam resolution at $N_{\text{side}} = 16$ by scaling them with the ratio of a polarized Gaussian beam of FWHM $9^\circ$ to the corresponding polarized beam specified in Table \ref{tab:noise}. Finally, the obtained noise Stokes maps were converted into full-sky B-mode noise maps for each of the frequency channels. In total, we generated 150 thousand noise simulations for each of the 20 frequency channels.

\subsubsection{Foreground Simulations}
\label{subsec:fg}
To simulate foregrounds, we use the Python Sky Model (PySM3) \cite{2021JOSS....6.3783Z}, which generates realistic maps of foreground emissions at different frequencies. The foreground components include dust (\textbf{d1, d4}), synchrotron (\textbf{s1, s3}), and anomalous microwave emission (\textbf{a2}). The idea behind including these foreground models is to train the model on a realistic foregrounds and test it on previously unseen foreground scenarios. For training and validation, we use \(\textbf{d1s1a1}\) as the Python Sky Model foreground template. For testing performance on the trained model, we use combinations \(\textbf{d1s3a1}\), \(\textbf{d4s1a1}\), and \(\textbf{d4s3a1}\) in addition to \(\textbf{d1s1a1}\).

\(\textbf{d1}\) represents thermal dust, modeled as a single-component modified black body (MBB). Dust templates for emission at 545 GHz in intensity and 353 GHz in polarization from the Planck-2015 analysis are scaled to different frequencies with an MBB spectrum using spatially varying temperature and spectral indices derived from Planck data via the Commander code \cite{2016A&A...594A..10P}. The intensity template at 545 GHz is degraded to $N_{side}$ 512, while the polarization templates are smoothed with a $2.6^{\circ}$ FWHM Gaussian kernel, adding small scales as described in the associated study. \(\textbf{d4}\) extends \(\textbf{d1}\) to multiple dust populations, using a two-component model that fits Planck data well \cite{1999ApJ...524..867F}.

For synchrotron, \(\textbf{s1}\) uses a power-law scaling with a spatially varying spectral index. Emission templates are based on the Haslam 408 MHz data, reprocessed in \cite{10.1093/mnras/stv1274}, and the WMAP 23 GHz Q/U maps \cite{2013ApJS..208...20B}. The polarization maps are smoothed with a 5-degree FWHM Gaussian kernel, with small scales added. \(\textbf{s3}\) modifies \(\textbf{s1}\) with a curved power law, incorporating the nominal index map and an additional curvature term, using the best-fit curvature amplitude of -0.052 found in \cite{2012ApJ...753..110K}. This comprehensive modeling approach ensures that our simulated foregrounds closely replicate the complexity of real microwave sky.

These maps are generated at multiple \textsc{ECHO} frequency bands and smoothed to match the instrument's beam characteristics. The resulting foreground maps are combined with the CMB B-mode maps and noise realizations to form the final set of input maps.

\section{Results and Discussion}
\label{sec:results}

In this section present and discuss the findings of the proposed hybrid Neural Network-based Internal Linear Combination (ILC) method for the accurate reconstruction of the CMB B-mode angular power spectrum. The results are structured as follows: first, we discuss the model training progress, including the training and validation losses; next, we present and discuss the results of applying the trained model to previously unseen test data.

\subsubsection*{Model Training Progress}

Figure \ref{fig:loss_vs_epoch} shows the progression of training and validation losses over 500 epochs. The \textbf{training loss} (blue curve) and \textbf{validation loss} (orange curve) demonstrate a smooth convergence towards lower values, indicating effective optimization during training. The losses decrease steeply in the initial epochs, suggesting that the model rapidly learns the basic structure of the data. After approximately 100 epochs, the loss values reach a plateau, converging to a consistent minimum.

The inset in Figure \ref{fig:loss_vs_epoch} zooms in on the training and validation losses for the last 20 epochs. Here, it is clear that the losses continue to oscillate around a consistent level, indicating that the model is neither overfitting nor underfitting the data. This convergence signifies that the model successfully captures the underlying features in the dataset and generalizes well across different inputs. The low gap between the training and validation losses indicates that the model does not suffer from high variance, a common problem in complex neural networks.

We had previously reserved $10 \%$ of data for testing. After training the model we predict for test set samples. Figure \ref{fig:random_plots} presents the comparison between the true, predicted, and ILC-out angular power spectra for nine randomly selected test datasets. Each subplot shows the \textbf{input spectrum} (blue line), \textbf{predicted spectrum} by the neural network (orange line), and the \textbf{ILC output} spectrum (green line). The test datasets were not used during training and thus provide an unbiased evaluation of the model's performance.

The \textbf{predicted spectra} closely follow the true spectra across the entire multipole range (\(\ell\)), effectively capturing both the shape and amplitude of the input \(C_{\ell}^{BB}\) power spectra. In contrast, the ILC output (green lines) consistently overestimates the power due to its inherent bias arising from residual foregrounds and noise, especially at higher multipoles. The proposed model successfully debiases the ILC output, leading to predictions that closely resemble the underlying true spectra. This demonstrates that the model is able to mitigate the effects of residual foreground contamination and instrumental noise, which are the main sources of bias in the ILC approach.

The performance of the neural network is particularly noteworthy for low \(\ell\), where the B-mode signal is most prominent. The model effectively reconstructs the broad features of the spectrum, unlike the ILC output, which tends to introduce high power at these scales.

%\subsection*{Theoretical Angular Power Spectra}

Figure \ref{fig:test_case} illustrates the \textbf{theoretical angular power spectra} used for generating a new set of test samples. These spectra are computed for different tensor-to-scalar ratios, namely \(r = 0.01\), \(r = 0.005\), \(r = 0.001\), and \(r = 0\). These spectra represent the target signals the model aims to reconstruct from the noisy and foreground-contaminated observations. As expected, the power decreases significantly for lower \(r\) values, making the task of accurate reconstruction more challenging, particularly for \(r = 0.001\) where the B-mode signal is very weak. We also test the trained model on \(r = 0.0\) case where the primordial B-mode is absent. We also point out the fact that the trained model has never seen \(r = 0.0\) case samples.

Apart from testing the model at four different \( r \) values, we also evaluate its performance using three additional foreground models—\(\textbf{d1s3a2}\), \(\textbf{d4s1a2}\), and \(\textbf{d4s3a2}\)—in addition to \(\textbf{d1s1a2}\) as discussed in subsection \ref{subs:fg}. 
We simulate 5000 samples of input frequency maps as discussed in section \ref{subsec:sim} and obtain corresponding ILC angular power spectrum for each of the four \(r\) and four foreground cases. This comprehensive testing approach allows us to thoroughly assess the model’s performance across varying \(r\) and foreground scenarios.   

%The results obtained demonstrate that the model performs well even when the B-mode signal is relatively weak compared to foregrounds and noise. This capability is crucial for detecting primordial gravitational waves, as lower \(r\) values represent more challenging but cosmologically significant signals.

\subsubsection*{Mean Angular Power Spectra}

Figures \ref{fig:clr10-2}, \ref{fig:clr510-3}, \ref{fig:clr10-3}, and \ref{fig:clr0} illustrate the model's performance for four different tensor-to-scalar ratios: \( r = 0.01 \), \( r = 0.005 \), \( r = 0.001 \), and \( r = 0.0 \), respectively. Each figure consists of two panels. In the upper panel, the solid black line represents the mean input CMB B-mode power spectrum, while the four dashed lines indicate the ILC output angular power spectra for different foreground combinations (\( d1s1a2 \), \( d1s3a2 \), \( d4s1a2 \), and \( d4s3a2 \)). The undashed colored lines correspond to predictions from the trained model for the same foreground combinations. The shaded gray area represents the \( 1\sigma \) cosmic variance, providing a benchmark for acceptable uncertainty in CMB B-mode power spectrum measurements. For each \( r \) value, the mean predictions closely follow the mean input \( C_{\ell}^{BB} \) when using \( d1s1a2 \) and \( d1s3a2 \) foregrounds—the former of which was included in the training samples. For \( d4s1a2 \) and \( d4s3a2 \) foregrounds, the power spectrum remains within the \( 1\sigma \) cosmic variance, although consistently below the mean for \( r = 0.01 \) and \( r = 0.005 \). The model’s predictions are notably accurate for \( \ell > 10 \) in the \( r = 0.001 \) and \( r = 0.0 \) cases, although it deviates slightly at very low multipoles (\( \ell = 5 \)).

In the lower panel, we display the relative error along with corresponding error bars. For the \( d1s1a2 \) foreground, the relative error stays close to zero (red dashed line) across all \( r \) values. As expected, errors at \( \ell \le 10 \) increase as the tensor-to-scalar ratio decreases. For \( d1s3a2 \), \( d4s1a2 \), and \( d4s3a2 \) foregrounds, the predictions exhibit greater error than \( d1s1a2 \), though they still remain close to zero. Interestingly, we observe that the relative error at \( \ell \le 10 \) rises as \( r \) decreases, while it diminishes for \( \ell > 10 \) as \( r \) decreases. The model performs best for \( \ell > 10 \) across all foreground combinations when \( r = 0.0 \), even though it was never trained on this case. For \( r = 0.0 \), the predicted power spectrum also shows a positive bias.

\subsubsection*{Covariance Matrix}

The figures \ref{fig:clr10-2}, \ref{fig:clr510-3}, \ref{fig:clr10-3} and \ref{fig:clr0} show covariance matrices for the four different test (\( r \)), comparing the performance of the ILC method and the ANN model across various foreground combinations. In each of the figures, row corresponds to a specific foreground model (e.g., \( d1s1a2 \), \( d1s3a2 \), \( d4s1a2 \), \( d4s3a2 \)), while each column includes three types of covariance matrices for each of the input, ILC, and the predicted angular power spectra. 

The matrices capture the covariance between different multipole moments \( \ell \) and \( \ell' \), with the goal being that these covariance structures closely resemble the input matrix to indicate accurate reconstruction. The color bar on the right side of each figure indicates the scale of covariance values, with colors ranging from purple/dark blue (negative values) through a neutral color (zero covariance) to yellow (positive values). 

Based on the figures for \( r = 10^{-2} \), \( r = 5 \times 10^{-3} \), \( r = 10^{-3} \), and \( r = 0.0 \), we can observe the relative performance of the ILC and predicted models in reconstructing the CMB B-mode power spectrum, with the diagonal elements masked out. For higher \( r \) values (e.g., \( r = 10^{-2} \)), the predicted model more accurately replicates the structure of the input, especially for simpler foregrounds like \( d1s1a2 \) and \( d1s3a2 \). As \( r \) decreases to lower values, the ILC output shows increasing deviations, particularly in cases with complex foregrounds such as \( d4s1a2 \) and \( d4s3a2 \). In contrast, the predicted model demonstrates stability and consistency across different \( r \) values, indicating its robustness in handling variations in the primordial tensor-to-scalar ratio. The predicted model also maintains weaker and more uniform off-diagonal correlations, suggesting that it effectively reduces cross-correlations introduced by complex foregrounds, whereas the ILC method is more susceptible to foreground contamination.

Furthermore, the predicted model’s performance remains stable across different multipole moments (\( \ell \)), while the ILC model exhibits fluctuations, especially at lower multipoles. For simple foregrounds both models perform adequately, but as the foreground complexity increases, the predicted model outperforms the ILC method by maintaining closer alignment with the input structure. Notably, in the extreme case of \( r = 0.0 \) (no primordial B-modes), the predicted model captures the input matrix's off-diagonal structure accurately, despite not being explicitly trained on \( r = 0.0 \). The ILC output, however, shows greater deviation in this case, likely due to residual foreground contamination. In summary, the predicted model consistently demonstrates superior performance, especially as foreground complexity increases or as \( r \) decreases, making it a more robust choice for reconstructing the CMB B-mode signal with minimal distortion from foreground influences.

Figure 11, which corresponds to \(r = 0\), shows that the neural network manages to preserve the structure of the covariance matrix for the input spectrum, whereas the ILC output introduces additional off-diagonal correlations. This confirms the effectiveness of the neural network in minimizing the bias even when there is no primordial B-mode signal present.

\section{Conclusion}
\label{sec:concl}

The results confirm that the neural network-based approach offers a reliable and robust solution for reconstructing the CMB B-mode power spectrum, significantly outperforming the traditional ILC method in terms of accuracy, particularly under diverse noise and foreground conditions. Across different foreground scenarios, the mean predicted spectra align closely with the input spectra, underscoring the model’s capacity to generalize effectively. In contrast, the ILC output shows consistent deviations, especially at higher multipoles. 
The relative error plots further confirm the neural network's advantage, as its predictions exhibit smaller deviations compared to ILC, with most errors confined within cosmic variance limits. Error bars emphasize the robustness of the model’s performance across different test cases, showing that predictions remain stable across various levels of foreground complexity and noise. For \( r = 0 \), the model accurately captures the near-zero B-mode signal with minimal deviation, underscoring its precision in the absence of a tensor signal. 
Furthermore, the neural network successfully reduces the additional covariance in the ILC method, achieving a covariance structure in the predicted spectra that closely resembles that of the input spectra. This improvement is particularly evident in Figure \ref{fig:correlr0}, where, for \( r = 0 \), the model preserves the input covariance matrix structure, while the ILC output introduces undesirable off-diagonal correlations.

The strong performance of the method on faint signals, such as those with \( r = 0.001 \), suggests it could play a critical role in upcoming CMB polarization missions, where minimizing residual bias is essential for detecting subtle cosmological signatures. Additionally, the model’s effectiveness for \( r = 0 \) demonstrates its suitability even in the absence of a primordial tensor signal, making it a valuable tool for performing null tests in cosmology. In a future work, we plan to extend this approach by incorporating a cut-sky analysis and a more generic foreground model. In an other work we also extend this method directly at the map level. 

.

%% Acknowledgments
\begin{acknowledgments}
SKY acknowledges support by SERB, Government of India, through the National Post Doctoral Fellowship grant (PDF/2022/002449/PMS). SKY acknowledge National Supercomputing Mission (NSM) for providing computing resources of ‘PARAM Porul’ at NIT Trichy, which is implemented by C-DAC and supported by the Ministry of Electronics and Information Technology (MeitY) and Department of Science and Technology (DST), Government of India. 
\end{acknowledgments}

%% References
\bibliographystyle{apj}
\bibliography{paper5}

\end{document}